\input{aipcheck}

\documentclass[
    ,final            
  ]
  {aipproc}

\layoutstyle{6x9}

\begin{document}

\title{Spin Physics at RHIC}

\author{L.C.Bland}{
  address={Brookhaven National Laboratory, Upton, NY   USA}
}

\begin{abstract}
 The physics goals that will be addressed by colliding polarized
 protons at the Relativistic Heavy Ion Collider (RHIC) are described.
 The RHIC spin program provides a new generation of
 experiments that will unfold the quark, anti-quark and gluon contributions
 to the proton's spin.  In addition to these longer term goals, this
 paper describes what was learned from the first polarized proton
 collisions at $\sqrt{s}$=200 GeV.  These collisions took place in a
 five-week run during the second year of RHIC operation.
\end{abstract}

\maketitle


\section{I.  Introduction}

  The Relativistic Heavy Ion Collider (RHIC) at Brookhaven National
  Laboratory was built to study the collisions of heavy ions having
  energies up to 100 GeV/nucleon.  The physics goal of these studies
  is to find a new state of matter where quarks and gluons are not
  confined in hadrons.  PHENIX and STAR are two large experiments and
  PHOBOS and BRAHMS are two small experiments
  designed to study relativisitic heavy ion collisions at RHIC.
  Details about these detectors can be found in
  Ref.~\cite{NIM_volume}.  

  In addition to having the capability to produce heavy
  ion collisions at energies up to $\sqrt{s_{NN}}=200$ GeV, RHIC is
  the first high-energy accelerator designed to allow the
  acceleration, storage and collisions of polarized proton beams
  with center of mass energies up to $\sqrt{s}=500$ GeV.  Helical dipole magnets
  in each ring serve as `Siberian Snakes' that help preserve the
  beam's polarization during acceleration and storage
  \cite{Huang}. A pair of Siberian Snake magnets in each ring make
  the vertical direction the stable spin axis.  These magnets were
  commissioned during RHIC run 2 to provide collisions of transversely
  polarized protons at $\sqrt{s}$=200 GeV.  Spin rotator magnets positioned on either side of
  the STAR and PHENIX interaction regions will permit future studies with
  longitudinally polarized proton beams.  Final preparations of the
  PHENIX and STAR detectors for the study of high-energy polarized
  proton collisions will soon be completed.  In addition, the PP2PP
  experiment is being built to study the elastic scattering of
  polarized proton beams at RHIC.

  This paper briefly describes the long-term physics goals of the RHIC
  spin program.  These goals have been described in a recent review \cite{Bunce}.  What is new
  in this area is the first collisions of polarized protons at RHIC.
  This paper provides a description of what was learned from the first
  RHIC run producing collisions of polarized protons.  Many
  details about the run were described in parallel session talks at
  SPIN 2002.  This overview serves as a roadmap to these more detailed
  descriptions.

\section{II.  Physics goals of the RHIC spin program}

  The RHIC spin program aims to unravel the spin structure of the
  proton.  Understanding how the valence quarks, gluons and sea
  quark-antiquark pairs conspire to produce the spin of the proton is
  equally as important as understanding how these constituents produce
  the proton's mass.  The quark contribution (${1 \over 2}\Delta
  \Sigma$), the gluon contribution ($\Delta G$) and possible
  orbital angular momentum of the proton's constituents must sum to
  the proton spin of ${1 \over 2}$.  To date, spin structure studies
  have relied on deep inelastic scattering (DIS) of charged leptons as
  was reviewed at this Symposium \cite{Averett,Miller}. The
  contribution quarks make to the overall spin of the proton has been
  found to be substantially smaller than expected
  ($\Delta\Sigma=0.23\pm0.04\pm0.06$).  Polarization of the gluons
  that bind the quarks and/or orbital angular momentum of the proton's
  constituents must carry the rest.  Since gluons carry no {\it
  electric} charge, real or virtual photons only probe the gluon
  through its splitting into $q\overline{q}$ pairs.  Use of this
  photon-gluon fusion process provides a means of establishing the gluon
  contribution to the proton's spin that is being actively pursued by the COMPASS
  experiment at CERN \cite{Tessarotto} and is a planned pursuit
  at SLAC \cite{Rock}.  COMPASS aims to probe gluon polarization by
  detecting $D$ mesons from $\gamma^* g\rightarrow c\overline{c}$
  interactions employing a polarized $\mu$ beam and a
  polarized target.  One of the goals of the RHIC spin
  program is to utilize the {\it color} charge of the quarks to couple
  directly to the gluon field within the proton.  Existing
  measurements of quark polarization from DIS experiments enable the
  use of quarks as an `analyzer' of gluon polarization in
  $\vec{p}\vec{p}$ collisions, particularly for final states where a
  large transverse momentum ($p_T$) photon is produced.  Gluon
  polarization will be probed at RHIC by measurements of longitudinal
  double spin asymmetries ($A_{LL}$).

  High energy collisions of protons produce hadronic jets,
  photons, vector bosons ($W^\pm$,$Z,\gamma^*$) and other particles having
  large $p_T$ or large mass.  From studies at unpolarized
  hadron colliders, it is known that perturbative QCD can provide
  quantitative predictions of the cross sections for such processes,
  particularly when next-to-leading order (NLO) contributions are included.
  Extending these phenomenlogical analyses to include spin degrees
  of freedom leads to the expectation that pQCD will provide a
  quantitative framework to interpret spin observables for jets,
  photons, and other particles produced in polarized proton collisions.

  As previously described \cite{Bland}, the `golden probe' of gluon
  polarization at RHIC will be photon production in longitudinally
  polarized proton collisions.  It is expected that the QCD Compton
  process ($qg \rightarrow q\gamma$) will be the dominant source of
  photons, with physics backgrounds arising from fragmentation photons
  and the $q\overline{q}\rightarrow\gamma g$ process, the latter
  process corresponding to time-reversed photon-gluon fusion.
  Experimentally, photon production is challenging because the total
  cross section is small relative to prolific sources of photons from
  neutral meson decays.  Simulations for both PHENIX and
  STAR have demonstrated that these backgrounds can be minimized.
  Coincident detection of the away-side jet with the prompt photon can
  provide information about the Bjorken $x$ values of the
  initial-state interacting partons.  Recent results from the Tevatron
  \cite{D0} suggest that NLO pQCD can provide
  a quantitative description of photon production cross sections.

  The production of $W^\pm$ bosons in the collision of longitudinally
  polarized protons is expected to result in spectacularly large
  parity violation.  As in charged current weak interactions that
  gives rise to the flavor sensitivity to neutrino deep inelastic
  scattering, parity violating spin asymmetries for $W^\pm$ production
  in $\vec{p}p$ collisions will
  isolate valence quark ($\Delta u/u,\Delta d/d$) and sea anti-quark
  ($\Delta \overline{u}/\overline{u}, \Delta \overline{d}/\overline{d}$) polarizations
  \cite{Saito}. Utilization of the electroweak interaction is expected
  to be a more precise measure of these polarizations than
  semi-inclusive DIS experiments that are
  presently underway.  Imprecise knowledge of fragmentation functions
  is a dominant systematic error for semi-inclusive DIS studies.

  The collision of transversely polarized proton beams can be used to
  study the transversity structure function.  Although this is a
  twist-2 function of equal importance as the longitudinal spin
  distribution functions ($\Delta q(x)$,$\Delta {\overline q}(x)$ and
  $\Delta G(x)$) for understanding the proton's spin structure,
  transversity cannot be probed in inclusive DIS because of helicity
  conservation.  Transverse spin asymmetry measurements at RHIC are
  expected to shed light on the transversity structure functions and
  should provide a timely complement to new measurements of transverse
  spin effects in semi-inclusive DIS \cite{Miller} and in jet studies
  from $e^+e^-$ collisions \cite{Hasuko}.

  The development of the RHIC accelerator complex to produce
  polarized proton collisions with a luminosity of $0.8 (2.0) \times
  10^{32}$ cm$^{-2}$s$^{-1}$ at $\sqrt{s}$=200(500) GeV and with beam
  polarizations of 70\% will be required to attain these long term
  physics goals.  Completion and full development of the PHENIX and
  STAR experiments is also required.  Intermediate physics goals can
  be attained in the first RHIC spin runs, providing an understanding of how to
  reach the requisite precision to measure small spin effects in the
  collider environment while providing the first glimpses of the
  proton's spin structure probed in high-energy polarized proton
  collisions.

\section{III.  Spin asymmetry measurements in a collider}

  A collider has significant differences, both advantageous and
  potentially problematic, from traditional fixed target
  experiments that severely impact the measurement of spin
  observables.  One positive aspect is that in a collider
  polarization reversals occur at the bunch crossing frequency.  For
  RHIC run 2, the inverse of this frequency was 214 nsec.  This is
  expected to be reduced to 107 nsec in subsequent runs.  These time
  scales are very short compared to most sources of time dependent
  variations in a detector's response, such as gain drifts, thereby
  reducing sensitivity to this class of systematic errors.  Rapid
  polarization reversals place stringent demands on timing of detector
  signals to preserve the correct association of events with the
  polarization state of the beams.  The detector response must also be
  minimal to backgrounds which are not associated with the passing beams.
  Out-of-time background events effectively obtain random polarization
  assignments.

  \begin{figure} 
  \caption{STAR data from a bunch crossing scaler system with input from a
  beam-beam counter coincidence \cite{Kiryluk} providing a measure of
  luminosity. There are two groups of 
  five bunch crossings with few counts, corresponding to the `abort
  gaps' (RF cycles with no beam) in the Blue and Yellow rings.  Counts
  in the abort gaps arise from single beam backgrounds.  The pattern of
  polarization directions for the Blue and Yellow beams at STAR is also shown.
  The non-uniformity of the yield versus bunch number produces 
  differences in the relative luminosity for different polarization
  directions.}
  \includegraphics[height=3.3in]{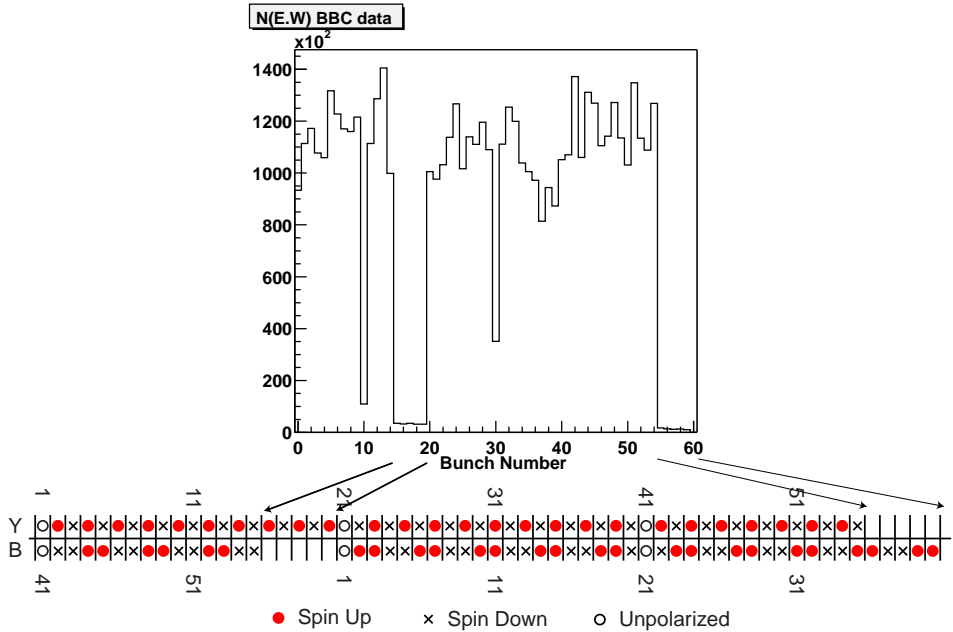}
  \label{pattern} 
  \end{figure}

  One significant challenge that is presented by collider experiments
  is the possibility of accidental correlation between polarization
  directions and luminosity of the colliding beams.  This correlation
  can arise because each ring is injected with individual bunches of
  protons with particular polarization orientations.  An injection
  pattern is chosen so that in one ring (Yellow) alternating bunches
  have opposite polarization direction and the other ring (Blue) has
  bunch pairs with alternating polarization (Fig.~\ref{pattern}).  Due
  to the potential correlation between luminosity and polarization,
  accurate measurement of the relative luminosity of bunch crossings
  with different polarization directions is required.  This is
  accomplished by counting experiments using fast detectors capable of
  discriminating collisions from beam-related backgrounds.  An example
  from RHIC run 2 is shown in Fig.~\ref{pattern}, with more details
  described in Ref.~\cite{Kiryluk}.  Methods are being developed to
  average out any relationship between intensity and polarization by
  reversing the spins of the stored beams and by recogging the beams
  so that different bunches collide.

  When measuring transverse single spin asymmetries, an azimuthally
  symmetric detector configuration can be used to measure
  spin-dependent cross ratios (spin up/down yields with left/right
  detectors), thereby making detector acceptance and luminosity
  asymmetries higher order corrections \cite{Spinka}. For observables
  involving longitudinal polarization there is no spin-dependent
  modulation of scattering rates that varies harmonically with the
  azimuthal angle.  Hence, accurate measurements must be made to
  prevent the relative luminosity determination from being a limiting
  systematic error of future $A_{LL}$ measurements.  The large cross section
  reactions needed for high rate relative luminosity monitors must
  themselves have minimal longitudinal double spin asymmetries.
  Identifying suitable monitoring reactions is one of the primary
  goals for RHIC run 3 when longitudinally polarized proton
  collisions are planned for the PHENIX and STAR experiments.

  It is planned to develop an AC dipole that will efficiently reverse
  the polarization of the stored beam \cite{Bai}.  This tool, used in conjunction
  with changes in which beam bunches overlap (recogging), will be fully
  developed in subsequent RHIC runs.  These developments are expected
  to further reduce spin dependent relative luminosity as a major
  source of systematic error for spin asymmetry measurements.

  Backgrounds in a collider also differ in important ways from those
  for fixed target experiments.  One difference between the two
  environments is that there are two beams in a collider, traveling in
  opposite directions, that can lead to twice the background.  As with
  the spin information, timing plays a critical role in discriminating
  signals from collisions versus those from backgrounds.  Time
  differences between counters mounted fore and aft of the interaction
  region can be used to identify collisions relative to beam-related
  backgrounds.

\section{IV.  First polarized proton collisions}

  The first collisions of transversely polarized protons occurred during RHIC
  run 2 in the period from early December, 2001 to late January,
  2002.  The first three weeks of this run were dedicated to accelerator
  commissioning, with a particular focus on commissioning the
  `Siberian Snake' magnet pairs in each ring.  Officially, data taking for $\sqrt{s}$=200 GeV
  $\vec{p}\vec{p}$ collisions began on December 20$^{th}$, initially at low
  luminosity.  Two weeks into the run, the average luminosity
  increased to $0.5\times10^{30}$ cm$^{-2}$ s$^{-1}$, and stayed at
  that value until the end of the run.  Knowledge of the beam polarization 
  is subject to the caveats discussed below.

  The RHIC polarimeters detect low-energy recoil carbon ions produced
  by $p + ^{12}$C elastic scattering reactions at small $|t|$, where
  the Coulomb scattering amplitude is comparable in magnitude to the
  strong-interaction amplitude.  Details about the design
  \cite{Kurita} and performance \cite{Jinnouchi} of these
  Coulomb-Nuclear Interference (CNI) polarimeters were described at
  this conference.  At present, the effective
  analyzing power of the CNI polarimeters is measured only near the
  injection energy \cite{Tojo}.  Ultimately, these polarimeters will
  be calibrated by determining the beam polarization from proton
  elastic scattering from a polarized hydrogen gas jet target,
  effectively transferring knowledge of the target polarization to
  provide knowledge of the beam polarization.  The first opportunity
  to make these measurements will occur during RHIC run 4.  Prior to
  that, the effective analyzing power of the CNI polarimeter can only
  be determined by special stores in RHIC involving an acceleration
  ramp followed by a deceleration ramp.  Measurements with the CNI
  polarimeter at the injection energy, flattop energy and then again
  at the injection energy can transfer the calibration of the
  polarimeter effective analyzing power to high energy, if there is
  found to be minimal polarization loss in both ramps.  Deceleration
  ramps were attempted during RHIC run 2, but were unsuccessful
  because of beam loss during deceleration.  Continuation of this
  development is planned for run 3.

  For RHIC run 2, knowledge of the magnitude of the beam polarization
  at the collision energy results from the assumption that the
  effective analyzing power of the CNI polarimeter does not depend on
  energy.  Model calculations suggest only a small increase in the
  effective analyzing power with increasing energy
  \cite{Trueman}. Based on the assumption that the effective analyzing
  power is independent of beam energy, the beam polarization at the
  collision energy averaged 17\% in the Yellow ring and 13\% in the
  Blue ring over the last two weeks of the run, with the difference
  reflecting subtle effects in the acceleration ramps of the two
  rings.  The ratio of the measured CNI polarimeter spin asymmetries
  at the collision energy to those at the injection energy for a given
  RHIC fill was on average smaller than unity.  Hence, the measured
  asymmetries at 100 GeV set an upper limit on the beam polarization
  at the collision energy, since acceleration is highly unlikely to
  increase the beam polarization.  The polarization magnitude was
  limited by the AGS, the last accelerator in the chain used to inject
  the RHIC rings.  Failure of the AGS power source necessitated use of
  a less powerful backup, resulting in a slower acceleration cycle,
  thereby reducing the polarization.

  \begin{figure}
  \centering
  \includegraphics[width=0.6\textwidth,bb=0 0 567 734,clip]{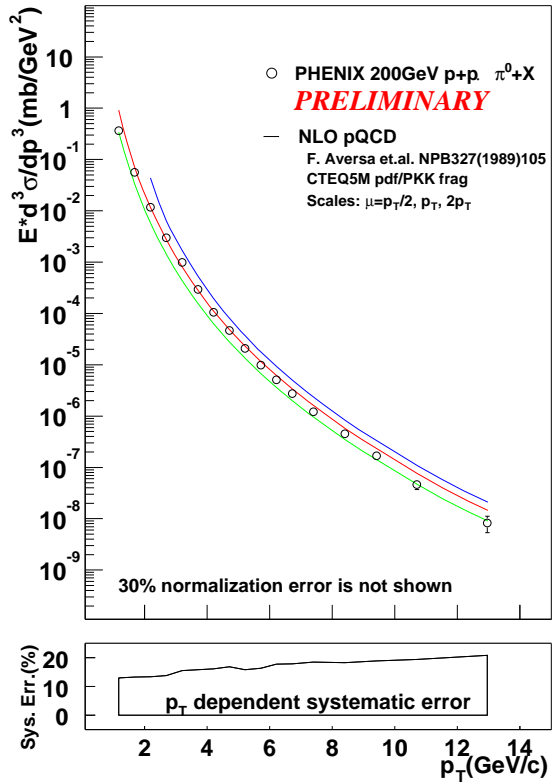}
  \caption{PHENIX results for mid-rapidity $p+p\rightarrow \pi^0+X$
  invariant cross sections versus $p_T$ measured at $\sqrt{s}=200$
  GeV in RHIC run 2.  The data are compared to NLO pQCD calculations.}
  \label{PHENIX_pi0}
  \end{figure}

  The physics goals of the first polarized proton collisions at
  $\sqrt{s}$=200 GeV included:
  \begin{itemize}
  \item{} the measurement of unpolarized observables to provide the
  requisite $pp$ reference data for the RHIC heavy-ion program,
  \item{} commissioning of the accelerator complex for polarized proton collisions,
  \item{} making the first measurements of spin asymmetries in a
  polarized proton collider,
  \item{} identifying reactions that will provide the basis for local
  polarimeters in subsequent runs which utilize the spin rotator
  magnets, and
  \item{} commissioning of the PP2PP Roman pots for the study of proton
  elastic scattering at small $|t|$.
  \end{itemize}
  These goals represent important benchmarks of progress to both the
  RHIC spin program and the RHIC heavy-ion program.  As is evident in
  the following description, these goals were met in the first
  polarized proton collision run.

\subsection{IV.a  Unpolarized proton observables}

  Many of the results for unpolarized $pp$ collision observables were
  reported at the recent Quark Matter 2002 conference \cite{QM02}.
  Minimum bias data samples were recorded at all RHIC experiments.
  High $p_T$ triggers based on the central arm electromagnetic
  calorimeters (EMC) enabled PHENIX to obtain the $p_T$ variation of
  mid-rapidity $\pi^0$ production out to 12 GeV/c
  (Fig.~\ref{PHENIX_pi0}).  These data are compared with
  next-to-leading order pQCD calculations, and agree with the
  calculations over an impressive eight orders of magnitude \cite{Fox}.  This
  good agreement between data and theory bodes well for using pQCD to
  interpret spin observables at RHIC energies.  PHENIX was also able
  to measure $J/\Psi$ production cross sections in their central
  detector arms by detecting $e^+e^-$ coincidences and in their south
  muon arm by detecting $\mu^+\mu^-$ coincidences \cite{Sato}.  Polarization
  observables for open charm production in $\vec{p}\vec{p}$ collisions
  is anticipated to be another important channel for studying gluon
  polarization.  PHENIX plans to commission their north muon arm
  during RHIC run 3, thereby completing the baseline detector.

  \begin{figure}
  \centering
  \includegraphics[height=2.0in]{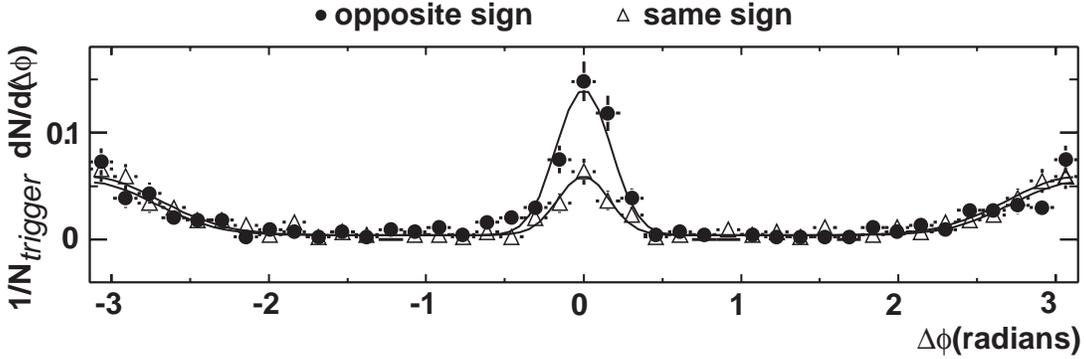}
  \caption{STAR results for azimuthal angle correlations between high
  $p_T$ charged hadrons produced in $p+p$ collisions at $\sqrt{s}$=200
  GeV \cite{Hardtke}.  The correlations are between a trigger particle with 4 $<
  p_T^{trig}<$ 6 GeV/c and associated particles with 2 GeV/c $<p_T<p_T^{trig}$.
  Both the near side ($\Delta\phi\sim0$) and away side
  ($|\Delta\phi|\sim\pi$) correlations are consistent with expectations
  from jet and dijet events.}
  \label{dihadron}
  \end{figure}

  STAR obtained a significant $pp$ minimum bias data sample that will
  provide critical reference data for its heavy-ion program.  The
  relatively low luminosity for the first polarized proton collisions
  minimizes the influence of `pile-up' background in the STAR time
  projection chamber (TPC).  In future years, higher luminosity $pp$
  collisions will require sophisticated offline analysis algorithms to
  discriminate charged particle tracks associated with the triggered event from
  pileup tracks that are produced in the $\sim$400 bunch crossings before or
  after the triggered event.  Even with only a minimum bias trigger,
  STAR embarked on its program of jet physics, exploiting the complete
  azimuthal coverage of the TPC for $|\eta|\le 1.4$.  As shown in
  Fig.~\ref{dihadron}, evidence for jet
  and di-jet events was obtained from di-hadron correlation studies \cite{Hardtke}.
  Cone and $k_T$ jet reconstruction alogrithms are presently being applied to
  the reconstructed TPC tracks for charged particles.  The results
  from these analyses are qualitatively consistent with expectations.
  Quantitative results for jet yields and correlations require
  detailed understanding of the TPC efficiency.  With the
  commissioning of the half barrel EMC
  ($2\pi$ azimuthal coverage for $0\le \eta \le 1$) in RHIC run 3, STAR
  will be able to trigger on and fully reconstruct jets.  Complete
  coverage from the barrel EMC ($-1\le \eta \le +1$), and the coverage from the
  endcap EMC ($1\le \eta \le 2$), will be added for subsequent runs.  Jet physics is an
  important component of the STAR spin physics program.

  \begin{figure}
  \caption{Results from PP2PP obtained during a single RHIC fill with
  reduced emittance and special focusing to permit positioning their
  Roman pots close to the beam.  The left (right) figure shows the
  horizontal (vertical) position correlation for events detected in
  silicon detectors mounted in the Blue and Yellow rings either side
  of the interaction point.  Elastic scattering events from $pp$
  collisions can be cleanly identified.}  
  \includegraphics[height=2.5in]{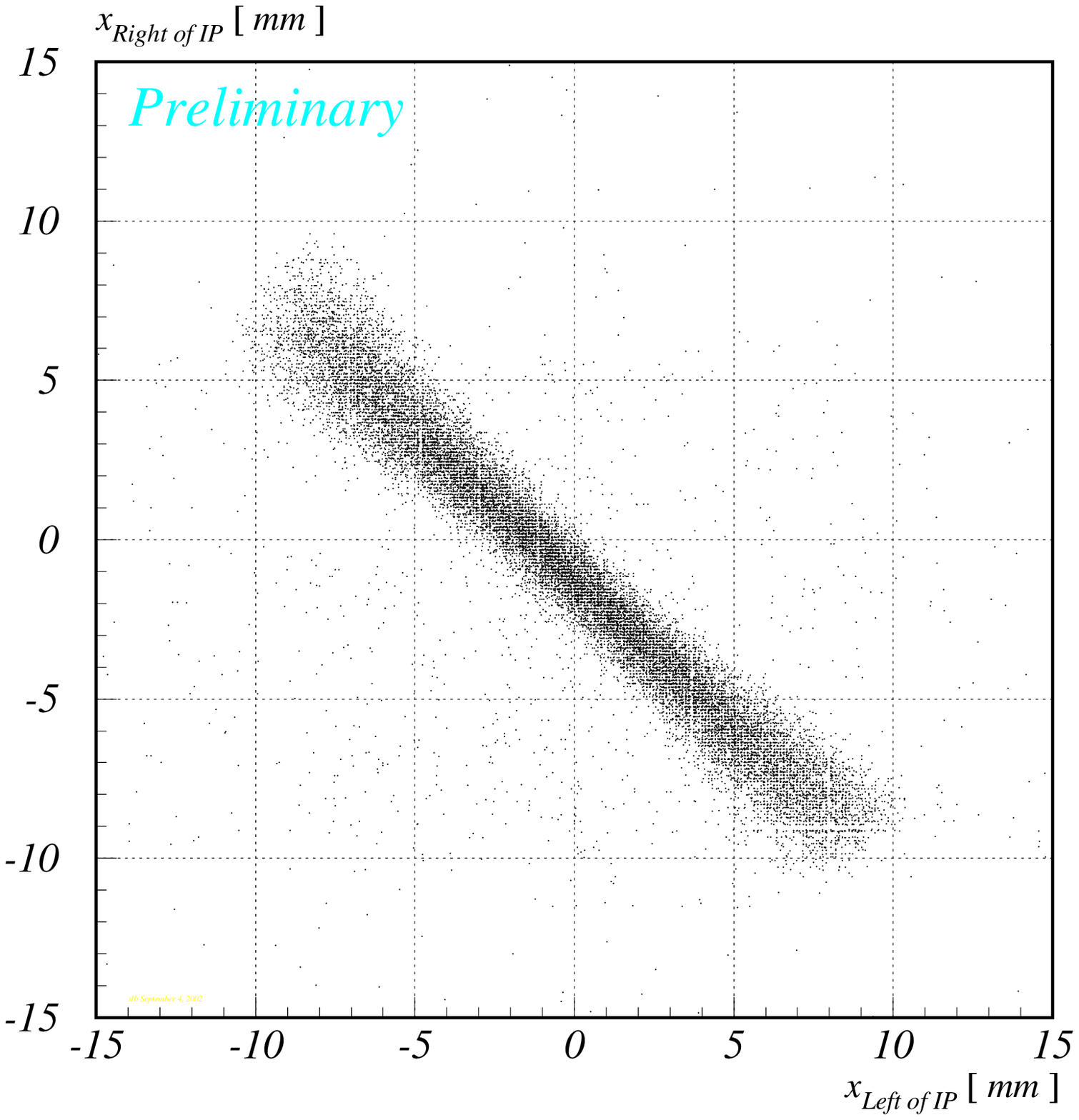}
  \hspace{1em}
  \includegraphics[height=2.5in]{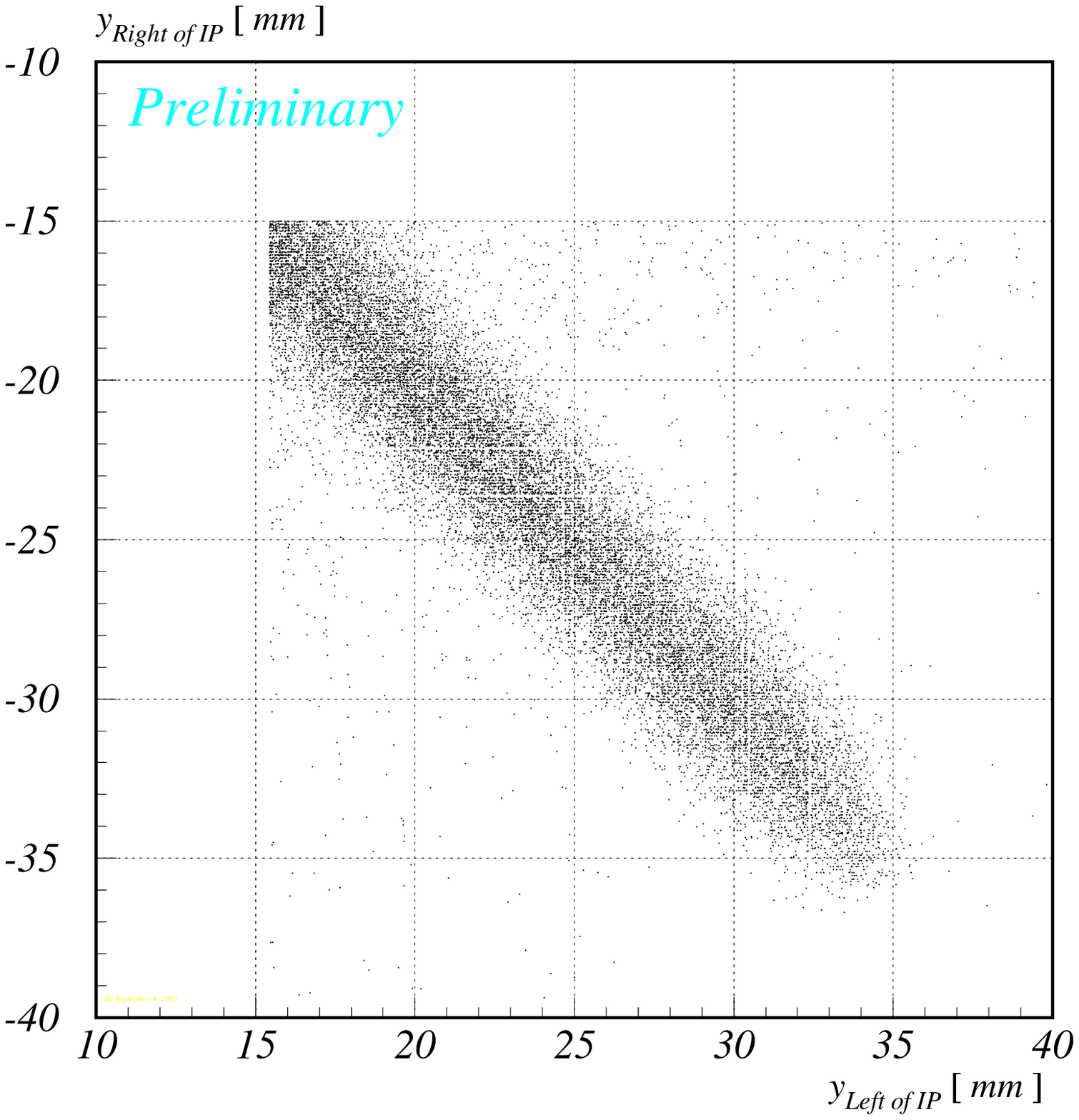}
  \label{pp2pp_data}
  \end{figure}

  RHIC run 2 also saw the commissioning of the PP2PP experiment.
  Their goal is to measure proton elastic scattering cross sections
  and spin observables for polarized proton collisions at
  $\sqrt{s}$=200 and 500 GeV.  The experiment utilizes `Roman pots' to
  place silicon detectors close to the beams, observing the
  elastically scattered protons in coincidence.  The emphasis is on
  measuring the $|t|$ dependence for elastic scattering in the
  vicinity of the maximal interference between the Coulomb and strong
  interaction amplitudes.  Elastic scattering events were observed
  \cite{Bueltmann} during the commissioning run in a dedicated store
  where the beams were scraped to reduce their emittance
  (Fig.~\ref{pp2pp_data}).  Further runs are planned in RHIC run 3, and
  beyond, to attain the goals of the physics program.

\subsection{IV.b  First spin asymmetry measurements and local polarimetry}  

  A suitable context for understanding the first polarized proton
  collisions at RHIC, which used transverse polarization, is the E-704 experiment at Fermi National
  Laboratory.  Prior to RHIC, E-704 studied polarized proton
  interactions at the highest $\sqrt{s}$.  E-704 was a fixed target
  experiment using a high-energy polarized proton beam produced by
  hyperon decay, resulting in polarized proton interactions at
  $\sqrt{s}$=20 GeV, an order of magnitude lower than the first RHIC
  spin run.  Large analyzing powers, increasing in magnitude as Feynman $x$
  increases beyond $x_F \sim 0.3$, were found by E-704 for both
  charged \cite{chargedpi} and neutral pion \cite{neutralpi} production at moderate $p_T$.  At
  mid-rapidity, analyzing powers were found to be consistent with
  zero.  Theoretical models that aimed to understand the E-704
  analyzing powers generally predicted similar trends at RHIC
  energies, even though they attributed the large Feynman $x$ analyzing
  powers to different dynamics.

  The E-704 results motivated two separate large $x_F$ measurements
  during RHIC run 2.  The goal of both of these measurements was to
  observe reactions with large analyzing powers that could provide the
  basis for local polarimeters at the PHENIX and STAR experiments.
  Spin rotator magnets will be used during RHIC run 3 to produce
  nominally longitudinal beam polarization at PHENIX and STAR.  Local
  polarimeters sensitive to either remnant vertical or radial
  polarization components are required to properly tune the spin
  rotator magnets.  Space constraints at PHENIX and STAR differ,
  implying that different techniques for local polarimetry are likely
  to be needed.

  \begin{figure}
  \caption{Results from the PHENIX spin group for the analyzing power
  for forward neutron production.  Shown in the figure are the
  measured spin-dependent asymmetries scaled by the Blue ring beam
  polarization, assuming the effective analyzing power of the CNI
  polarimeter does not depend on energy.}
  \includegraphics[height=2.5in]{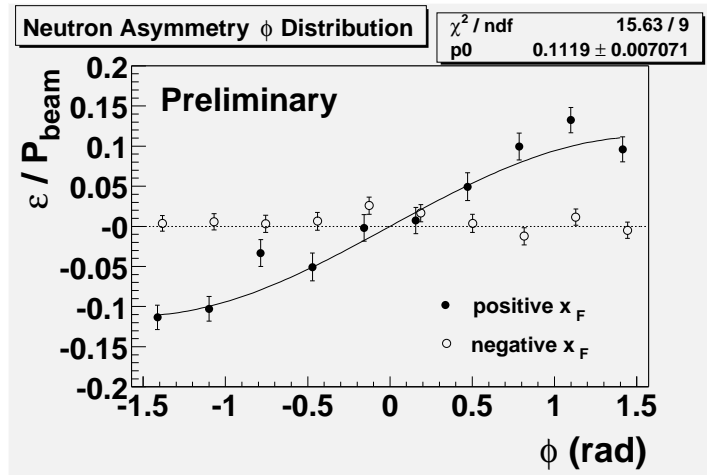}
  \label{neutron_asymmetry}
  \end{figure}

  One experiment was conducted at interaction point 12 (IP12) in the
  RHIC ring by members of the PHENIX spin group \cite{Fukao}. They
  designed and constructed a calorimeter consisting of a 5
  (horizontal) $\times$ 12 (vertical) matrix of PbWO$_4$ detectors
  that was positioned 18 meters distant from the interaction point
  immediately following the DX magnet east of IP12, and hence
  sensitive to spin effects from collisions associated with the
  polarization direction of the proton beam in the
  Blue ring.  The objective was to be sensitive to possible analyzing
  powers associated with large $x_F$ $\gamma$ or $\pi^0$ production.
  The geometry of their calorimeter implied that 50 GeV energy
  deposition corresponded to a maximum $p_T$ of 150 MeV/c.  The
  experiment could discriminate incident $\gamma$ and neutrons by use
  of counters that were located in front of and behind the calorimeter.  Through the
  course of the run, a portion of a Zero Degree neutron Calorimeter
  (ZDC) module was installed west of IP12, again 18 meters distant
  from the interaction point, to confirm results that were obtained
  from analysis of the data during the run.

  The conclusion from this experiment (Fig.~\ref{neutron_asymmetry}) is
  that there are sizable analyzing powers for low $p_T$ neutron
  production in polarized proton collisions at $\sqrt{s}$=200 GeV.
  These results will serve as the basis of local polarimetry at the
  PHENIX experiment in RHIC run 3 by equipping the ZDC's fore and aft
  of PHENIX with position sensitive shower maximum detectors.

  At STAR, electromagnetic calorimeters were mounted upstream of the
  DX magnet in close proximity to the beam pipe and 7.5 meters distant
  from the STAR interaction point \cite{Rakness}. A Pb-scintillator
  sampling calorimeter, equipped with a two orthogonal planes of
  scintillator strips serving as a position-sensitive shower maximum
  detector was mounted north of the beam.  Simple $4\times4$ matrices
  of lead glass detectors were mounted in three locations, directly
  south of and above and below the beam pipe.  Since the calorimeters
  were mounted east of STAR, they were sensitive to spin effects from collisions
  associated with the polarization direction of the proton beam in the Yellow ring.  The calorimeter
  mounted north of the beam was able to identify neutral pions with
  good invariant mass resolution, and was sensitive to transverse
  momenta in the range, $1 \le p_T \le 4$ GeV/c.

  \begin{figure}
  \includegraphics[width=3.0in]{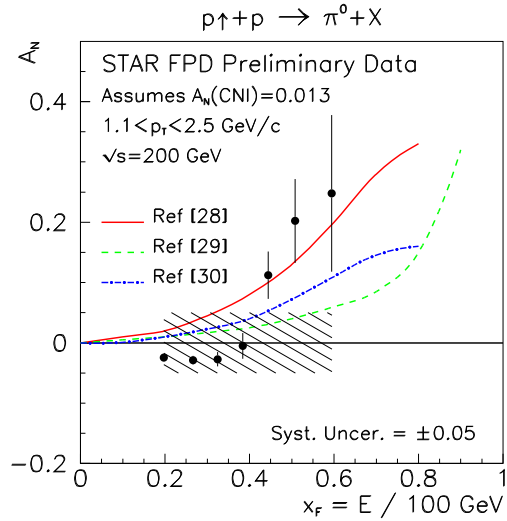}
  \caption{STAR results for the analyzing power for events dominated
  by $\pi^0$ production at large pseudorapidity ($3.4\le \eta \le
  4.1$).  The data are compared to calculations available prior to the
  measurement.  Ref.~\cite{Collins} is based on the Collins effect,
  corresponding to a spin dependent fragmentation.  Ref.~\cite{Qiu} is
  based on a twist-3 quark-gluon correlation responsible for the spin
  effect evaluated at $p_T$=1.5 GeV/c.  Ref.~\cite{Sivers} is based on
  the Sivers effect, where the spin effects arise from a correlation
  between the quark spin and its transverse momentum in the
  distribution function.  } 
  \label{pi0}
  \end{figure}

  The conclusion from the STAR measurement is that forward neutral
  pion production at RHIC energies has large analyzing powers that
  increase with increasing Feynman $x$.  The results at $\sqrt{s}$=200
  GeV (Fig.~\ref{pi0}) bear a strong similarity to the E-704 results
  \cite{neutralpi} and to theoretical predictions based on the lower
  energy data.  Given that the effective analyzing power of the CNI
  polarimeter is measured only at the RHIC injection energy, these
  results are {\it lower limits} on the $\pi^0$ analyzing power at
  $\sqrt{s}$=200 GeV.  Attempts are underway to equip STAR with
  electromagnetic calorimetry mounted at large pseudorapidity as a
  more permanent Forward $\pi^0$ Detector (FPD).  The new FPD will
  enable further study of the Feynman $x$ and $p_T$ dependence of
  analyzing powers for forward $\pi^0$ production, and can also serve
  as a local polarimeter for the tuning of the spin rotator magnets.

  In addition to these two experiments aimed at identifying local
  polarimeters for the future, analyzing powers for particle
  production at mid-rapidity will result from RHIC run 2.  STAR
  reported the analyzing power for leading charged particles at
  mid-rapidity out to $p_T$=5 GeV/c \cite{Balewski}. As expected by
  naive pQCD, $A_N$ is found to be zero at mid-rapidity.  PHENIX will
  ultimately be able to make quantitative statements to much higher
  $p_T$ from their mid-rapidity neutral pion and charged hadron data.

  In summary, the polarized proton run at $\sqrt{s}$=200 GeV during
  RHIC run 2 accomplished important goals, including:
  \begin{itemize}
  \item{} measurement of mid-rapidity particle production cross
  sections over a broad range of $p_T$.  These measurements are well
  described by NLO pQCD, which bodes well for future RHIC studies of
  the spin structure of the proton,
  \item{} the clear observation of jet events in $pp$ collisions.  Jet production and
  $\gamma$+jet coincidences for longitudinally polarized proton
  collisions will be a primary means of establishing gluon
  polarization,
  \item{} the observation of $pp$ elastic scattering
  collisions.  Only small backgrounds are observed in silicon detectors placed close
  to the beams, thereby demonstrating that RHIC is a well controlled
  accelerator providing a clean environment for studing $pp$ elastic
  scattering for colliding beams in the CNI
  regime (small $|t|$), and
  \item{} the observation of large spin asymmetries in the collision
  of transversely polarized protons.  These measurements demonstrate
  proper handling of the spin labeling of bunches and matching with
  the RHIC polarimeter results.  The reactions that give rise to these
  large analyzing powers will be used as local polarimeters at PHENIX
  and STAR in future runs, and themselves represent important
  transverse spin physics results.
  \end{itemize}
  Low polarization from the AGS is an important issue that must be
  addressed in future RHIC spin runs.  New polarimeter and accelerator
  hardware are expected to increase the polarization of beams stored in RHIC.

\section{V.  Plans for future runs}

  Spin rotator magnets have now been installed at
  the PHENIX and STAR experiments and will be commissioned during RHIC
  run 3.  These magnets will precess the
  proton beam's polarization from the vertical stable spin direction
  set by the two helical dipole Siberian Snakes in each ring, to
  longitudinal at the interaction points of these two experiments, and
  then back again.  Tuning the spin rotator magnets requires robust
  local polarimetry to establish that vertical and radial
  polarization components are small.  The analyzing power results from
  RHIC run 2 should provide the requisite feedback for this important
  development work.  In addition to this development, the luminosity
  for polarized proton collisions in RHIC run 3 is projected to be
  $10^{31}$ cm$^{-2}$s$^{-1}$ by using a tighter focus ($\beta^*$=1 m)
  and by increasing the number of bunches in the Blue and Yellow ring
  from 55 to 110.  The high-power generator is expected to be repaired
  and used for the AGS.  The projected beam polarization for RHIC run
  3 is 40\%, with a goal of further improvements on that value.
  Polarized proton collisions will be at $\sqrt{s}$=200 GeV for run 3.

  \begin{figure}
  \caption{Simulations for longitudinal spin correlations ($A_{LL}$)
  for (left) inclusive particle production at PHENIX using the
  polarized parton distributions from Ref.~\cite{Gehrmann} and (right)
  inclusive jet production at STAR for $0\le \eta \le 1$ using polarized parton
  distributions from Ref.~\cite{GRSV}.  All $2\rightarrow2$ subprocesses are included in
  the simulations.  The STAR simulations account for trigger and
  reconstruction biases.}
  \includegraphics[height=2.1in]{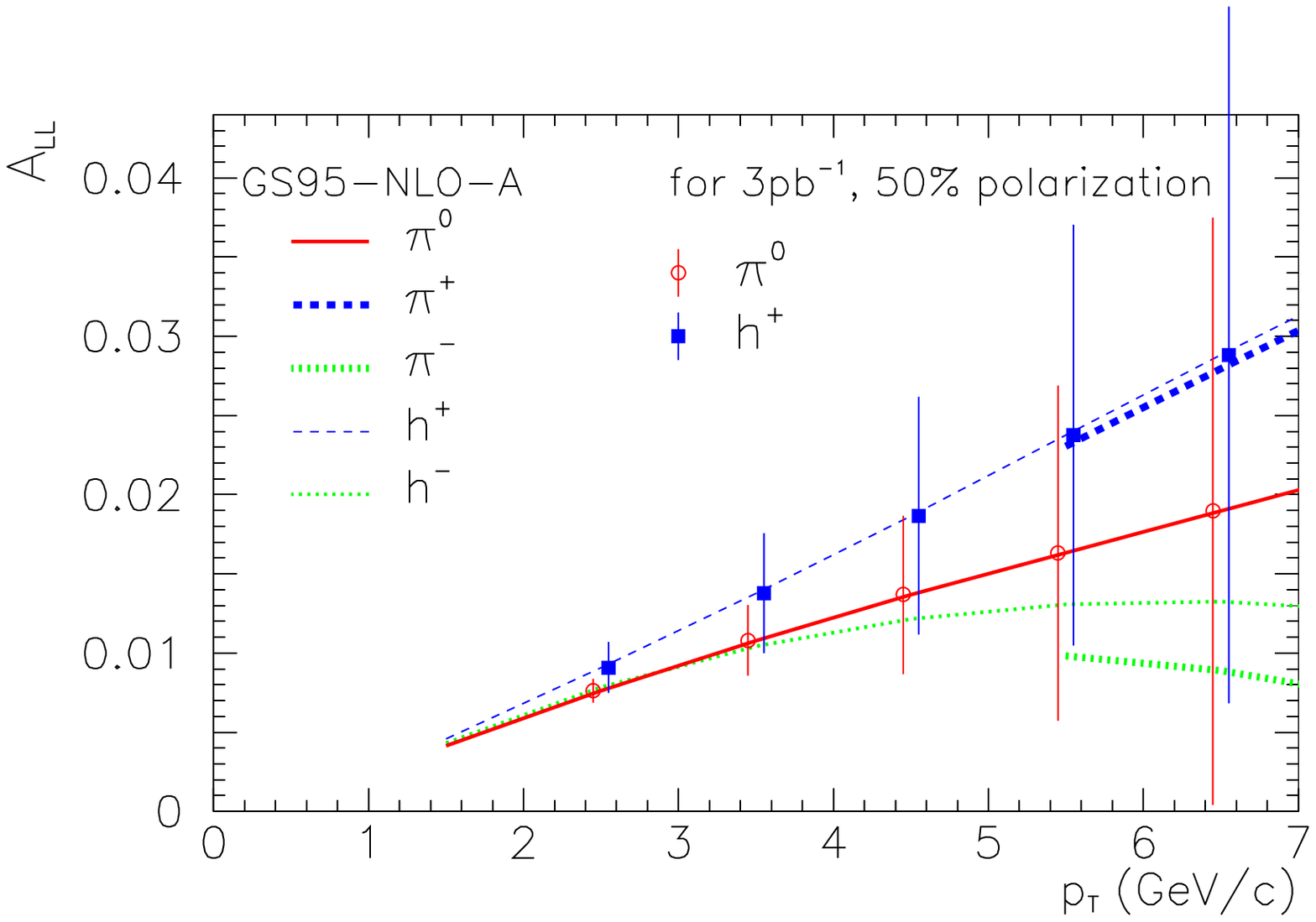}
  \hspace{1em}
  \includegraphics[height=2.4in]{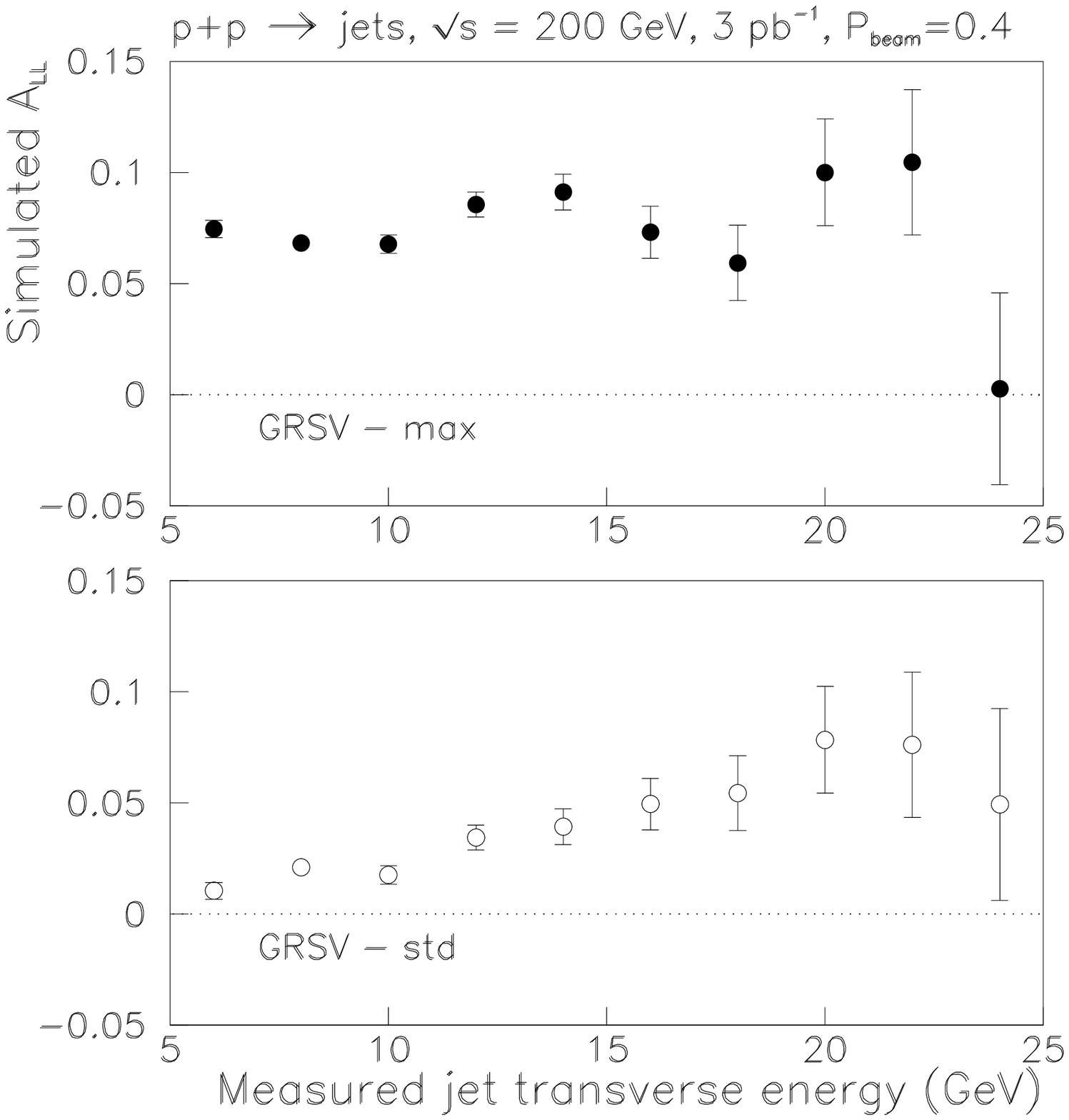}
  \label{jet_all}
  \end{figure}

  With longitudinal polarization for both proton beams, and the
  projected improvement in the polarized beam operation of the AGS
  injector to RHIC, it should be possible to embark on the first
  measurements of $A_{LL}$ for mid-rapidity $\pi^0$ production at
  PHENIX and for mid-rapidity ($0\le \eta \le 1$) jet production at STAR
  (Fig.~\ref{jet_all}).  Projections for the integrated luminosity and
  the beam polarization for polarized beam operations in RHIC run 3
  suggest sufficient statistical sensitivity to discriminate extreme
  scenarios for the degree to which gluons are polarized within the proton.
  Systematic errors need to be sufficiently controlled to provide
  robust measurements of $A_{LL}$.  Careful relative luminosity measurements
  for bunch crossings with equal and opposite longitudinal
  polarizations are required as is an absolute calibration of the
  analyzing power of the RHIC polarimeter for proton beam energy of 100 GeV.

  Looking beyond RHIC run 3, there are several accelerator and
  detector developments on the horizon:

  \begin{itemize}
  \item{} Polarized beams will be accelerated to 250 GeV to allow
  collisions at $\sqrt{s}$=500 GeV.  The highest collision energies
  are important for determining gluon polarization at small $x$ and
  necessary for studying spin asymmetries for $W^\pm$ production.  
  \item{} The polarized hydrogen gas jet target is expected to be
  first available for RHIC run 4.  This will ultimately provide a
  robust means of determining the beam polarization with a projected
  accuracy of 5\% expected to be attained during RHIC run 5.
  \item{} Improvements in the AGS polarization are expected for runs 4
  and 5, leading to the goal of 70\% polarization for the beams stored
  in RHIC.  A new fast polarimeter, based on the RHIC CNI polarimeter,
  has been built for the AGS for run 3, and several new hardware
  improvements are being considered for the AGS for runs 4 and 5.
  These include a strong partial Siberian Snake, expected to be
  available for run 5.
  \item{}  The STAR barrel and endcap electromagnetic calorimeters
  will be completed and commissioned for RHIC run 5.  These
  calorimeters provide $2\pi$ azimuthal coverage for the
  pseudorapidity interval $-1\le \eta \le 2$, playing a crucial role
  in spin physics measurements involving jets, photons and high-energy
  electrons and positrons.
  \item{}  Upgrades to PHENIX are planned to improve triggering
  capabilities for the muon arms required to fully exploit their
  $W^\pm$ physics program. There are also plans to improve their
  mid-rapidity charged particle tracking.  The latter will enhance jet
  reconstruction capabilities at PHENIX.   
  \end{itemize}

  The success of the first polarized proton collision run upholds the
  promise that the RHIC spin program will provide important new
  insight to the spin structure of the proton.

\begin{theacknowledgments}

  This report summarizes the work of many people who are part of the
  RHIC spin collaboration (RSC).  The RSC has members from the RHIC
  accelerator groups, the PHENIX collaboration, the STAR
  collaboration, the PP2PP experiment and the theory community that
  frequently communicate and exchange important information about
  polarized proton interactions at RHIC.  RIKEN
  has played a critical role in supporting the development of the hardware
  for polarized proton collisions at RHIC and the creation of the
  RIKEN/BNL Research Center, a stimulating environment for
  exploring RHIC physics.  The US Department of Energy and the
  National Science Foundation are playing central roles in supporting
  the RHIC spin program.

\end{theacknowledgments}


\bibliographystyle{aipproc}   


\IfFileExists{\jobname.bbl}{}
 {\typeout{}
  \typeout{******************************************}
  \typeout{** Please run "bibtex \jobname" to optain}
  \typeout{** the bibliography and then re-run LaTeX}
  \typeout{** twice to fix the references!}
  \typeout{******************************************}
  \typeout{}
 }

\end{document}